\pdfoutput=1
\documentclass[aps,prl,twocolumn, titlepage,showpacs]{revtex4}

\usepackage{graphicx}
\usepackage{color}
\usepackage{dcolumn}

\usepackage{bm}

\begin{document}

\title{Longitudinal viscosity of 2D Yukawa liquids}

\author{Yan Feng}
\email{yanfengui@gmail.com}
\altaffiliation{Present address: Los Alamos National Laboratory, Mail Stop E526, Los Alamos, New Mexico 87545, USA.}
\author{J. Goree}
\author{Bin Liu}
\affiliation{Department of Physics and Astronomy, The University
of Iowa, Iowa City, Iowa 52242}

\date{\today}

\begin{abstract}

The longitudinal viscosity $\eta_l$ is obtained for a two-dimensional (2D) liquid using a Green-Kubo method with a molecular dynamics simulation. The interparticle potential used has the Debye-H\"{u}ckel or Yukawa form, which models a 2D dusty plasma. The longitudinal $\eta_l$ and shear $\eta_s$ viscosities are found to have values that match very closely, with only negligible differences for the entire range of temperatures that is considered. The bulk viscosity $\eta_b$ is determined to be either negligibly small or not a meaningful transport coefficient, for a 2D Yukawa liquid.

\end{abstract}

\narrowtext

\maketitle

\section{I.~Introduction}
The longitudinal viscosity $\eta_l$ is a transport coefficient of interest for fluids~\cite{Sanchez:85}. It is the counterpart to the better-known transverse viscosity~\cite{Bernu:78}, which is more commonly called the shear viscosity $\eta_s$. The latter characterizes the momentum flux perpendicular to a velocity gradient. These viscosities are theoretically predicted to be related by~\cite{Vollmayr:11, Garzo:11}
\begin{equation}\label{lviscosity}
{\eta_l = 2 \frac{d-1}{d} \eta_s + \eta_b,}
\end{equation}
where $d$ is the dimensionality of the system, and $\eta_b$ is the bulk viscosity. The bulk viscosity $\eta_b$~\cite{Dukhin:09} is also called the volume viscosity or expansive viscosity; it is a parameter for liquids as well as molecular gases~\cite{bulkvis}. Both the shear and bulk viscosities appear in the Navier-Stokes equation~\cite{Landau:87} of a fluid. However, as compared with the shear viscosity $\eta_s$, the longitudinal viscosity $\eta_l$ and the bulk viscosity $\eta_b$ are studied less often.

Physically, these kinds of viscosity characterize energy dissipation in a fluid. Bulk viscosity is for energy dissipation due to compression and rarefaction of a fluid, for example in shock waves and high-frequency sound waves. Shear viscosity, on the other hand, is for energy dissipation due to a gradient in the flow velocity. In the latter case, the energy dissipation rate is proportional to both the shear viscosity and the square of the velocity gradient~\cite{Feng:12, Feng:12_1}. In the case of a periodic density perturbation, the energy dissipation rate is proportional to both the bulk viscosity and the square of the rate of density change~\cite{Sawyer:89, Madsen:92}.

Unlike shear viscosity, longitudinal and bulk viscosities are in general difficult to measure experimentally~\cite{Malbrunot:83, Dukhin:09}. This is because the hydrodynamic effects of bulk viscosity are significant only for rapid time variations, unlike shear viscosity which affects flows in easily observed ways, even under steady conditions. Ultrasound attenuation has been described as the only experimental method available for measuring bulk viscosity of a fluid~\cite{Malbrunot:83, Dukhin:09}. In this method, one can obtain the bulk viscosity after subtracting the ultrasound attenuation contributions from thermal conduction and shear viscosity~\cite{Prangsma:73, Malbrunot:83, Dukhin:09}. This method of measuring bulk viscosity is so difficult that it has been used only for water and a handful of exotic liquids~\cite{Dukhin:09}, and even for these substances the results for the bulk viscosity have large uncertainties. In contrast to this difficult experimental situation, however, longitudinal viscosity and bulk viscosity are mentioned more often in the theoretical literature, where it has been calculated for example using the Green-Kubo relation~\cite{Bernu:77, Bernu:78, Hoover:80, Hoheisel:86, Hoheisel:87, Tankeshwar:96, Hess:01, Okumura:02, Salin:02, Salin:03, Fernandez:04, Okumura:04, Meier:05, Bastea:07, Palla:08, Baidakov:11}, using the hydrodynamic limit~\cite{Vieillefosse:75}, or derived using a Chapman-Enskog approach~\cite{Chang:64}. In this paper, we will make use of the Green-Kubo approach, which we present in Sec.~II. We will use the Green-Kubo method to obtain $\eta_l$ and $\eta_s$, and for comparison we will use Eq.~(\ref{lviscosity}) to study $\eta_b$.

Dusty plasma~\cite{Melzer:08, Morfill:09, Piel:10, Shukla:02, Bonitz:10} is partially ionized gas containing micron-size solid particles, also called dust particles. These dust particles are highly charged negatively within the plasma by absorbing more electrons than ions, since negatively charged electrons have a higher temperature than positively charged ions. Due to the shielding provided by free electrons and ions in the plasma, the interaction between dust particles in a plasma can be modeled using a Yukawa or Debye-H\"{u}ckel potential~\cite{Konopka:00}, similar to charged particles in a colloidal suspension~\cite{Lowen:93}. Because of the high particle charge, dust particles in plasmas are strongly coupled (i.e., the potential energy between neighboring particles is larger than its kinetic energy), so that the collection of dust particles exhibits properties of liquids or solids. In laboratory experiments, dust particles can be in two-dimensional (2D) or three-dimensional (3D) suspensions, depending on the experimental conditions. In 2D experiments, all dust particles are confined in a horizontal plane, with negligible out-of-plane motion due to strong confining potentials in the vertical direction. The dust particles are immersed in a rarefied gas, which applies a much weaker friction to moving dust particles, as compared with the case of a colloid. The size of dust particles allows imaging them directly and tracking their motion, so that various transport mechanisms can be studied experimentally at the particle level. Transport mechanisms that have been studied for dusty plasmas include diffusion~\cite{Liu:08}, shear viscosity~\cite{Nosenko:04}, and thermal conduction~\cite{Nosenko:08}. For 2D dusty plasmas, viscosity is generally attributed to dust particle scattering arising from interparticle interactions, while scattering due to the molecules of rarefied gas is negligible, as explained in~\cite{Feng:11}. Shear viscosity has been widely studied for dusty plasmas, first in simulations for 3D systems~\cite{Sanbonmatsu:01, Saigo:02, Salin:02}, then later in experiments~\cite{Nosenko:04, Feng:11} and simulations~\cite{Liu:05, Donko:06} for 2D systems, as well as 3D experiments~\cite{Gavrikov:05, Vorona:07}. The longitudinal viscosity and the bulk viscosity have been quantified for classical 3D one-component plasmas (OCP)~\cite{Bernu:77, Bernu:78, Salin:02, Salin:03} using molecular dynamics (MD) simulations and gluon plasmas~\cite{nuclearplasma}, however, it has until now not been quantified for classical 2D dusty plasmas, to the best of our knowledge.

In this paper, we will report a determination of the longitudinal viscosity for a 2D Yukawa liquid. We will also report an unusual finding regarding the bulk viscosity: it is either negligibly small or it is not a meaningful transport coefficient, for a 2D Yukawa liquid.

\section{II.~Green-Kubo relations for $\eta_l$ and $\eta_b$}
Green-Kubo relations are often used to calculate various transport coefficients, such as diffusion~\cite{Vaulina:08}, shear viscosity~\cite{Liu:05, Feng:11}, and thermal conductivity~\cite{Donko:09}. Green-Kubo relations are for equilibrium conditions; they use microscopic random motion of particles to determine transport coefficients without any macroscopic gradients. The longitudinal and bulk viscosities can also be calculated using the Green-Kubo relations~\cite{Bernu:77, Bernu:78, Hoover:80, Hoheisel:86, Hoheisel:87, Tankeshwar:96, Hess:01, Okumura:02, Salin:02, Salin:03, Fernandez:04, Okumura:04, Meier:05, Bastea:07, Palla:08, Baidakov:11} using similar equations as the shear viscosity. The required inputs for calculating longitudinal and bulk viscosities include time series of particles' positions, velocities, and interparticle forces.

Now we review the three steps for calculating the longitudinal viscosity using the standard Green-Kubo relation~\cite{Bernu:78, Hoover:80, Hoheisel:86, Hoheisel:87, Tankeshwar:96, Hess:01, Okumura:02, Salin:03, Fernandez:04, Okumura:04, Meier:05, Bastea:07, Palla:08, Baidakov:11}. These Green-Kubo relations, which were originally developed for three dimensions $(d = 3)$, are adapted here for two dimensions $(d = 2)$ by setting the velocity and coordinate in the $z$ direction to be zero.

First, we calculate a diagonal element of the stress tensor $P_{xx}(t)$, which is defined as
\begin{equation}\label{SS}
{P_{xx}(t)=
\sum_{i=1}^N\left[mv_{ix}v_{ix}-\frac{1}{2}\sum_{j\not=i}^N\frac{x_{ij}x_{ij}}{r_{ij}}\frac{\partial \Phi(r_{ij})}{\partial r_{ij}}\right].}
\end{equation}
Here $i$ and $j$ indicate different particles which all have the same mass $m$, $N$ is the total number of particles, $\mathbf{r}_{i} = (x_i,y_i)$ is the position of particle $i$, $x_{ij}=x_i-x_j$, $y_{ij}=y_i-y_j$, $r_{ij}=|\mathbf{r}_i-\mathbf{r}_j|$, and $\Phi(r_{ij})$ is the interparticle potential energy. The positions and velocities of particles in Eq.~(\ref{SS}) vary with time, and this accounts for the time dependence of $P_{xx}(t)$. The off-diagonal element of the stress tensor, $P_{xy}(t)$, can be used to calculate the shear viscosity~\cite{Feng:11, Liu:05}. Unlike $P_{xy}(t)$ which fluctuates around zero, however, $P_{xx}(t)$ fluctuates around a constant level $\overline {P_{xx}(t)}$.

Second, we calculate an autocorrelation function for the fluctuation of $P_{xx}(t)$ using
\begin{equation}\label{SACF}
{C_l(t)= \langle (P_{xx}(t) - \overline {P_{xx}(t)}) (P_{xx}(0) - \overline {P_{xx}(t)}) \rangle.}
\end{equation}
Here $C_l(t)$ is the stress autocorrelation function. The brackets $\langle \cdot\cdot\cdot \rangle$ indicate an average over an equilibrium ensemble, which in practice we replace by an average over different initial conditions.

Third, we integrate the stress autocorrelation function over time to yield the longitudinal viscosity $\eta_l$~\cite{Hansen:86, Tankeshwar:96}
\begin{equation}\label{etal}
{\eta_l=\frac{1}{A k_B T}\int^\infty_0 C_l(t)dt,}
\end{equation}
where $A$ is the area of the 2D system and $T$ is its temperature. (For a 3D system, $A$ would be replaced by the system volume $V$.) These equations represent the Green-Kubo relation for the longitudinal viscosity in 2D systems. To improve statistics, we calculate $\eta_l$ twice, using $P_{xx}$ as shown above and also using $P_{yy}$, and we average the resulting values of $\eta_l$.

In addition to the longitudinal viscosity, we can also calculate the shear viscosity $\eta_s$~\cite{Liu:05, Donko:09, Feng:11} for 2D systems using
\begin{equation}\label{Pxy}
{P_{xy}(t)=
\sum_{i=1}^N\left[mv_{ix}v_{iy}-\frac{1}{2}\sum_{j\not=i}^N\frac{x_{ij}y_{ij}}{r_{ij}}\frac{\partial \Phi(r_{ij})}{\partial r_{ij}}\right],}
\end{equation}
\begin{equation}\label{SACFs}
{C_s(t)= \langle P_{xy}(t) P_{xy}(0) \rangle,}
\end{equation}
and
\begin{equation}\label{etas}
{\eta_s=\frac{1}{A k_B T}\int^\infty_0 C_s(t)dt.}
\end{equation}
The bulk viscosity~\cite{Bernu:78, Hoheisel:86, Tankeshwar:96, Salin:02} for 2D systems can be calculated similarly, using
\begin{equation}\label{AverageP}
{\widetilde{P(t)}=\frac{1}{2}(P_{xx}(t)- \overline {P_{xx}(t)}+P_{yy}(t) -\overline{P_{yy}(t)}),}
\end{equation}
\begin{equation}\label{SACFb}
{C_b(t)= \langle \widetilde{P(t)} \widetilde{P(0)} \rangle,}
\end{equation}
and
\begin{equation}\label{etab}
{\eta_b=\frac{1}{A k_B T}\int^\infty_0 C_b(t)dt.}
\end{equation}
We will use the same simulation data as the inputs in calculations of $\eta_l$ and $\eta_s$.

It has been questioned theoretically whether transport coefficients are meaningful for 2D liquids. This question has been studied theoretically, starting with a 2D hard disk system~\cite{Ernst:70} and then liquids with other interparticle potentials~\cite{Donko:09}. A transport coefficient is deemed to be not meaningful if the corresponding autocorrelation function has a long-time tail that decays as slowly as $1/t$, so that the Green-Kubo integral does not converge. For a 2D Yukawa liquid, the validity of transport coefficients has been discussed in detail in~\cite{Liu:05, Feng:11, Ott:08, Donko:09, Ott:09}. In Sec.~IV, we will present our autocorrelation functions and discuss whether they have a long-time tail.

Equations~(\ref{SS}-\ref{etab}) are presented in physical units, although we will perform simulations using dimensionless units. Some of the parameters we will use when making quantities dimensionless include the area $A$ of the simulated system, the areal number density $n$, the particle mass $m$, the Wigner-Seitz radius $a \equiv (n\pi)^{-1/2}$, a characteristic plasma frequency~\cite{Kalman:04} $\omega_{pd} = (Q^2/2\pi\epsilon_0ma^3)^{1/2}$, and the particle kinetic temperature $T$. Here, $Q$ is the particle charge.

\section{III.~Simulation method}
To model 2D dusty plasmas, we perform equilibrium MD simulations using a binary interparticle interaction with a Yukawa potential~\cite{Konopka:00}. We integrate the equation of motion $m\ddot{\mathbf{r}}_{i}=-\nabla \sum \phi_{ij}$ for all particles. This equation of motion does not include any friction term or any Langevin heating term. Particles are constrained to move only within a single 2D plane. Our simulation includes $N = 1024$ particles in a rectangular box with periodic boundary conditions to model an infinite system. The Yukawa potential is $\phi_{ij} = Q^2 {\rm exp}(-r_{ij}/\lambda_D)/(4\pi\epsilon_0 r_{ij})$, where $\lambda_D$ is the screening length. We truncate the Yukawa potential at distances beyond a cutoff radius of $24.76~a$; this truncation has been justified in~\cite{Liu:05}. This simulated system is essentially the same as a Yukawa OCP, except that we constrain the particle to move only on a single plane at $z = 0$.

Yukawa systems can be described by two dimensionless parameters: the coupling parameter $\Gamma$ and the screening parameter $\kappa$. They are defined as $\Gamma=Q^2/(4\pi\epsilon_0ak_BT)$ and $\kappa\equiv a/\lambda_D$. One can think of $\Gamma$ as an inverse temperature and $\kappa$ as an inverse indicator of density.

The input parameters in our simulation include $\kappa$ and $\Gamma$. We choose a single value of $\kappa = 0.5$, which is typical for 2D dusty plasma experiments~\cite{Feng:11}. When $\kappa = 0.5$, the melting point of 2D Yukawa system is $\Gamma \approx 142$~\cite{Hartman:05}. To study 2D Yukawa liquids over a large temperature range, we choose twelve different values of $\Gamma$ varying from 140 (corresponding to a temperature near the melting point) down to 2 (corresponding to a much higher temperature). The integration time step is in a range of $0.0037$ and $0.037~\omega_{pd}^{-1}$, depending on the choice of $\Gamma$, as in~\cite{Liu:05}. For each value of $\Gamma$, we perform four runs with different initial configurations of particles.

\begin{figure}[htb]
\centering
\includegraphics{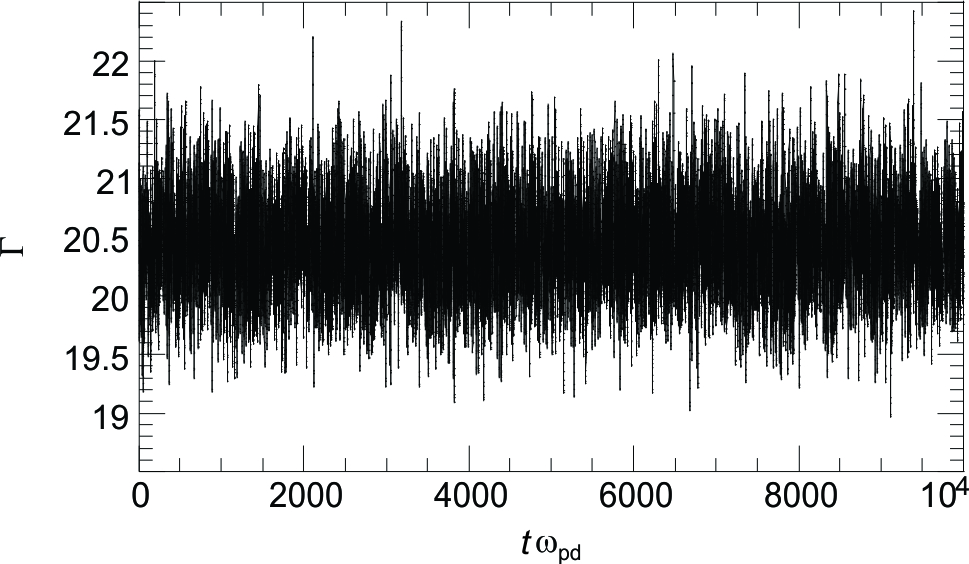}
\caption{\label{temperature} The temperature fluctuation during about one half of the $10^6$ steps used for data analysis for $\Gamma = 20$ and $\kappa = 0.5$. Here, $\Gamma^{-1}$ is a dimensionless temperature.}
\end{figure}

We use a thermostat only for the initial equilibrium of our simulation, and not for the data used to calculate $\eta_l$ and $\eta_s$. For each simulation run, we first integrate $10^5$ steps using a Nos\'e-Hoover thermostat to approach equilibrium at a desired temperature~\cite{Liu:05} under steady conditions. We then turn off this thermostat to integrate another $10^6$ steps. Only the data in the latter $10^6$ steps will be used to calculate the viscosities. We use a sufficiently small time step so that the energy conservation is adequately obeyed during the simulation run, as we have verified for our simulation data. We measure $T$, which can differ slightly from the desired temperature, using the mean square velocity fluctuation.

Figure~1 shows the time series of the measured temperature from one of our simulation runs. The temperature fluctuates about a steady level during the $10^6$ steps simulation interval. The temperature fluctuations are due to the finite simulation size. The absence of a general upward or downward trend in the temperature as a function of time is due to our choice of an adequately small integration time step. When we report a value for $\Gamma$, we use a temperature that was averaged over the $10^6$ steps, for a given run.

\section{IV.~Results}

\subsection{A.~Longitudinal and shear viscosities}

\begin{figure}[htb]
\centering
\includegraphics{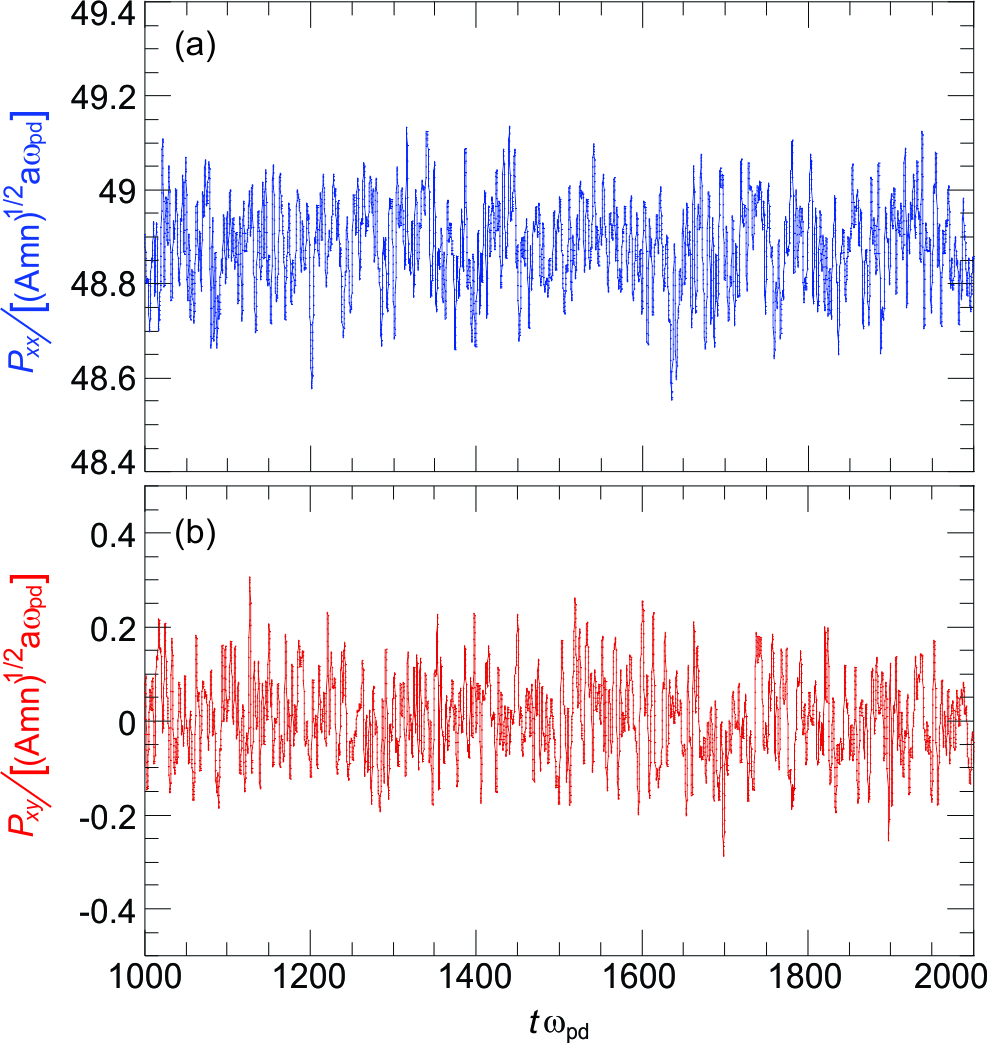}
\caption{\label{stress} (Color online). The fluctuations of the stress tensor elements during about $5\%$ of the $10^6$ steps used for data analysis for $\Gamma = 20$ and $\kappa = 0.5$.}
\end{figure}

Using the particles' positions, velocities, and potentials from the simulation, we use Eqs.~(\ref{SS}) and (\ref{Pxy}) to calculate the time series of the stress tensor elements. Examples of the results for the stress tensor are shown in Fig.~2. All our calculations of $\eta_l$ and $\eta_s$ will be based on these time series. We note that $P_{xx}(t)$ fluctuates about a non-zero value, while $P_{xy}(t)$ fluctuates about zero. The source of this fluctuation is the microscopic compression and shear motion of particles. We find that fluctuations of $P_{xx}(t)$ and $P_{xy}(t)$ have comparable amplitudes and time scales.

\begin{figure}[htb]
\centering
\includegraphics{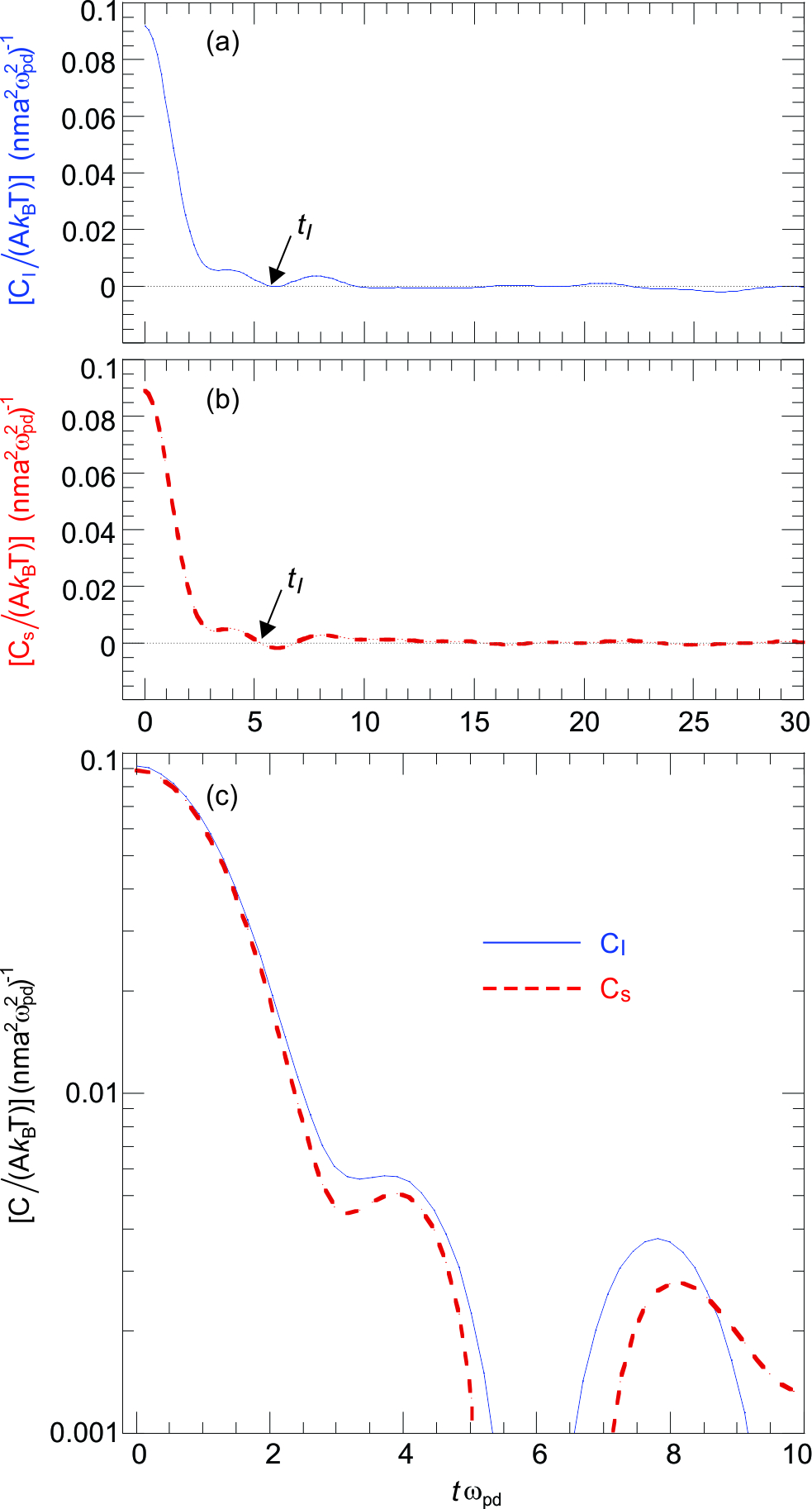}
\caption{\label{SACFf} (Color online). The autocorrelation functions of the stress tensor elements, where $C_l$ for longitudinal viscosity and $C_s$ for shear viscosity, for $\Gamma = 20$ and $\kappa = 0.5$. The same autocorrelation functions are shown with linear axes in (a) and (b), and a log axis in (c). Data are normalized by parameters defined in Sec.~II and III. The first zero crossings in (a) and (b) are marked as $t_I$, which will be used to replace the upper limits in the integrals, Eqs.~(\ref{etal}) and (\ref{etas}).}
\end{figure}

We calculate the autocorrelation function of the time series of $P_{xx}(t)$ using Eq.~(\ref{SACF}), and the result is shown in Fig.~3. Also shown is the autocorrelation function of $P_{xy}(t)$. In panels (a-b), we see the initial decay followed by a small variation around zero. We will use the area under these curves to calculate the viscosities, using Eq.~(\ref{etal}) for $\eta_l$ and Eq.~(\ref{etas}) for $\eta_s$. The integral in Eq.~(\ref{etal}) has a finite upper time limit of infinity, which we replace with the first zero crossing~\cite{Feng:11}, marked as $t_I$ in Fig.~3. In panel (c) we show the positive portion of these autocorrelation functions on a logarithmic scale so that the two curves can be distinguished. The two curves are almost identical in the initial decay, which leads us to expect that the longitudinal and shear viscosities will have almost the same values.

\begin{figure}[htb]
\centering
\includegraphics{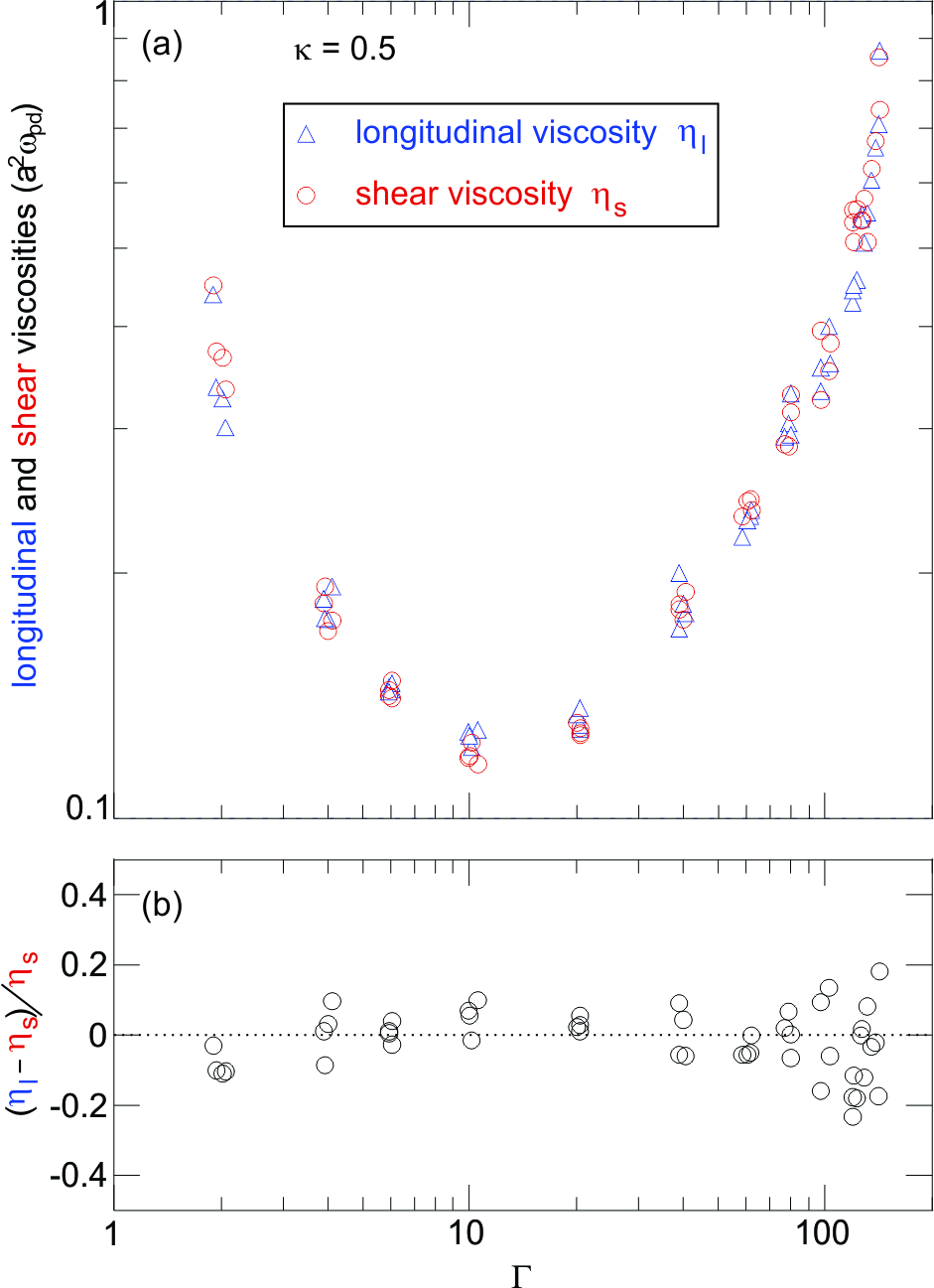}
\caption{\label{etasim} (Color online). (a) The longitudinal and shear viscosities from evaluating Eq.~(\ref{etal}) and Eq.~(\ref{etas}), respectively. (b) The difference between $\eta_l$ and $\eta_s$ is negligible, for all values of $\Gamma$. Using Eq.~(\ref{lviscosity2D}), this result indicates that $\eta_b$ is negligibly small. The minimum in $\eta_s$ in (a) is known to arise from a balance of kinetic and potential terms in the shear stress~\cite{Liu:05, Donko:06}, and here we find that a similar minimum appears in $\eta_l$.}
\end{figure}

Indeed, we find that the longitudinal viscosity $\eta_l$ and the shear viscosity $\eta_s$ have almost the same values, for the full range of $\Gamma$ that we investigate. This is seen in Fig.~4(a), where we present $\eta_l$ and $\eta_s$ determined by performing the integrals of the stress autocorrelation function in Eqs.~(\ref{etal}) and (\ref{etas}). We find negligible differences between them, as shown in Fig.~4(b).

We also see a minimum in $\eta_l$ as a function of $\Gamma$. This minimum matches the minimum in $\eta_s$, which was previously studied and explained as being due to a balance of kinetic and potential terms in the shear stress~\cite{Liu:05, Donko:06}.

In Fig.~4, the data have scatter in both axes. For the horizontal axis, the scatter around each value of $\Gamma$ indicates slight differences in the measured temperature, which can occur due to the absence of a thermostat, as discussed in Sec.~III. For the vertical axis, the scatter corresponds to the random run-to-run variation for the obtained viscosity values, i.e., random errors. In addition to these random errors, there is a systematic error associated with the choice of the upper integral limit. By examining the fluctuation of the Green-Kubo integral at long times~\cite{Danel:12}, we determined that this systematic error is smaller than the random errors.

\subsection{B.~Bulk viscosity}

The negligible difference between $\eta_l$ and $\eta_s$ that we find in Fig.~4 leads us to determine that the bulk viscosity $\eta_b$ is much smaller than either $\eta_l$ or $\eta_s$. This conclusion is drawn from Eq.~(\ref{lviscosity}) for our 2D system, which is
\begin{equation}\label{lviscosity2D}
{\eta_b = \eta_l - \eta_s.}
\end{equation}

Previous simulations using the Green-Kubo approach to obtain the bulk viscosity were mostly for 3D Lennard-Jones interparticle potentials and soft-sphere interparticle potentials~\cite{Hoover:80, Hoheisel:86, Hoheisel:87, Tankeshwar:96, Okumura:02, Fernandez:04, Okumura:04, Meier:05, Bastea:07, Palla:08, Baidakov:11}, or similar interparticle potentials~\cite{Hess:01}. In some of those simulations, the shear viscosity was also calculated~\cite{Hoover:80, Hoheisel:86, Hoheisel:87, Fernandez:04, Okumura:04, Baidakov:11}, and it was found that the bulk viscosity differs from the shear viscosity, with the difference within one order of magnitude.

Simulations of 3D plasmas~\cite{Bernu:77, Bernu:78, Salin:02, Salin:03} provided results for the bulk viscosity for both one-component plasmas (OCP) and Yukawa OCP. In these simulations~\cite{Bernu:77, Bernu:78, Salin:02, Salin:03}, it was found that the bulk viscosity is negligible as compared with the shear viscosity, about two orders of magnitude smaller or even more. From this aspect, it seems that our results for 2D Yukawa system that $\eta_b \ll \eta_s$ are consistent with those previous simulations in 3D OCP systems.

\begin{figure*}[htb]
\centering
\includegraphics{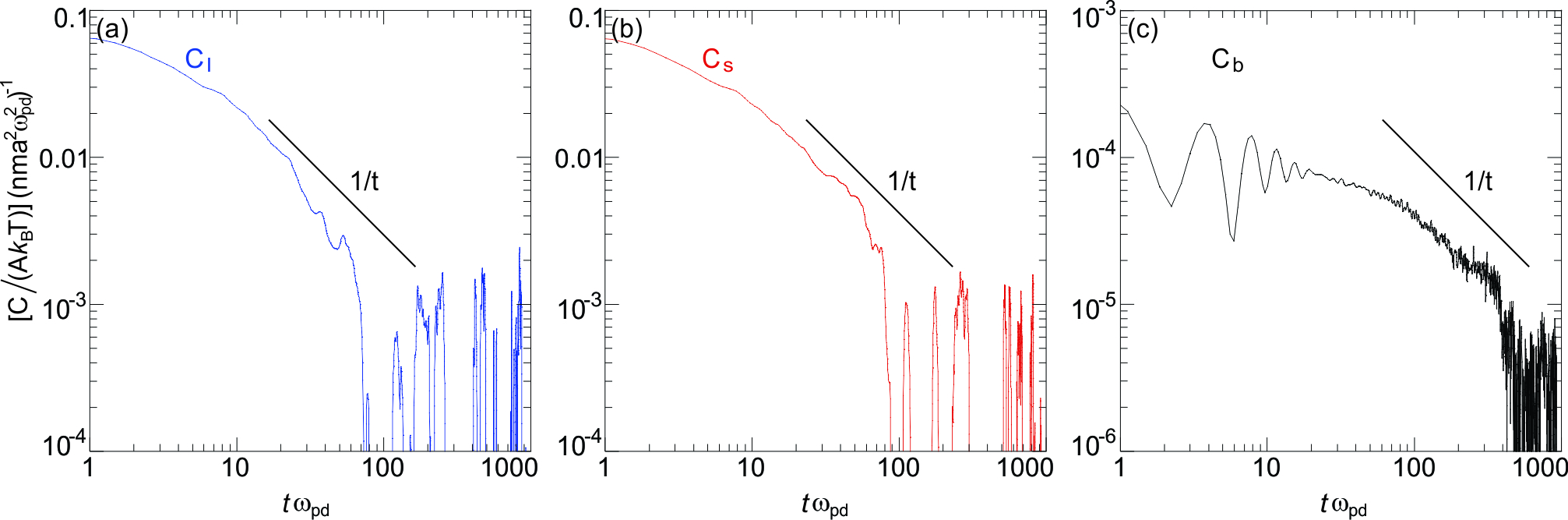}
\caption{\label{longtimetail} (Color online). Correlation functions for (a) $\eta_l$, (b) $\eta_s$, and (c) $\eta_b$, for $\Gamma = 140$ and $\kappa = 0.5$. Here, data are shown with log-log axes to allow an identification of any possible long-time tail. We find that only the correlation function $C_b$ for the bulk viscosity in (c) has a significant long-time tail, as seen by a decay that is slower than $1/t$. These results, for $\Gamma = 140$, are representative of the other values of $\Gamma$ studied as well.}
\end{figure*}

We now examine the autocorrelation functions used in calculating $\eta_l$, $\eta_s$, and $\eta_b$ in Eqs.~(\ref{etal}, \ref{etas} and \ref{etab}) to determine whether they exhibit a long-time tail. As discussed in Sec.~II, if the correlation function decays more slowly than $1/t$, this long-time tail prevents the convergence of the Green-Kubo integral so that the corresponding transport coefficient is deemed to be not meaningful. In Fig.~5(a-b) we present the correlation functions $C_l$ for longitudinal viscosity and $C_s$ for shear viscosity, and we find that they do not exhibit a noticeable long-time tail before the function becomes noisy. However, in Fig.~5(c) the correlation function $C_b$ decays more slowly, as can be seen by comparing to the line drawn with a slope corresponding to a $1/t$ scaling. This result suggests that, within the uncertainties that are inherent in a finite-size simulation~\cite{Donko:09}, $\eta_l$ and $\eta_s$ are meaningful, but $\eta_b$ is not. It is interesting that the signal-to-noise ratio for $C_b$, the correlation function of the bulk viscosity, is still comparable to that of $C_s$ and $C_l$, even though its amplitude is one or two orders of magnitude smaller. Even if $\eta_b$ were meaningful, it would have a small value because $C_b$ in Fig.~5(c) is two orders of magnitude smaller than $C_l$ and $C_s$.

We cannot explain in terms of the {\it macroscopic} fluid equations why the bulk viscosity is either negligibly small or not meaningful for this 2D liquid. However, in terms of {\it microscopic} motion, we can discuss some of the terms of the correlation functions. The correlation function for the longitudinal viscosity $C_l$ involves only products of $P_{xx}$ with a delayed version of itself, and likewise for $P_{yy}$. The bulk viscosity has a different character because Eq.~(\ref{SACFb}) also includes cross terms like $\langle \widetilde{P_{xx}(t)} \widetilde{P_{yy}(0)} \rangle$. In fact, the correlation function for the bulk viscosity, Eq.~(\ref{SACFb}), can be written as the sum of two terms: $C_l/2$ and a cross correlation involving $P_{xx}$ and $P_{yy}$. For our 2D Yukawa liquid these two terms almost cancel.

\section{V.~Summary}

Molecular-dynamics simulations of a 2D Yukawa liquid demonstrate that the longitudinal viscosity is almost the same as the shear viscosity, over a wide range of temperature. These results were obtained using Green-Kubo integrals of the appropriate autocorrelation functions. The very close match of the values for $\eta_l$ and $\eta_s$ would predict, using Eq.~(\ref{lviscosity}), that $\eta_b$ is negligibly small or even zero. Indeed, $\eta_b$ might not even be a meaningful transport coefficient for the system studied here because we found that its autocorrelation function exhibits a long-time tail, so that the corresponding Green-Kubo integral diverges. This divergence does not occur for $\eta_s$ and $\eta_l$, as they do not have a long-time tail, as judged by our simulation. We note that our results are based on a finite-size simulation; future larger simulations might be able to provide noise-free correlation-function data for longer times to further test these conclusions.

This work was supported by NSF and NASA.

\end{document}